\documentclass[10pt,superscriptaddress,tightenlines,twocolumn,amsmath,amssymb,aps, pra]{revtex4-2}

\usepackage{amsmath,braket,bm}
\usepackage{graphicx}
\usepackage{dcolumn}
\usepackage{bm}

\usepackage{color}

\usepackage[ruled]{algorithm2e}
\usepackage{newfloat,algcompatible}
\usepackage{physics}

\usepackage{booktabs} 
\usepackage{tabularx}

\newcommand{\ttt}[1]{\texttt{#1}}

\newcounter{rowItemCount}%
\newcounter{subRowItemCount}%

\begin{document}

\title{Neural network-based prediction of the secret-key rate of quantum key distribution}

\author{Min-Gang Zhou}
\author{Zhi-Ping Liu}
\author{Wen-Bo Liu}
\affiliation{National Laboratory of Solid State Microstructures, School of Physics, Collaborative Innovation Center of Advanced Microstructures, Nanjing University, Nanjing, China}
\author{Chen-Long Li}
\author{Jun-Lin Bai}
\author{Yi-Ran Xue}
\affiliation{National Laboratory of Solid State Microstructures, School of Physics, Collaborative Innovation Center of Advanced Microstructures, Nanjing University, Nanjing, China}
\affiliation{MatricTime Digital Technology Co. Ltd., Nanjing, China}
\author{Yao Fu}
\affiliation{MatricTime Digital Technology Co. Ltd., Nanjing, China}
\author{Hua-Lei Yin}
\affiliation{National Laboratory of Solid State Microstructures, School of Physics, Collaborative Innovation Center of Advanced Microstructures, Nanjing University, Nanjing, China}
\author{Zeng-Bing Chen}
\affiliation{National Laboratory of Solid State Microstructures, School of Physics, Collaborative Innovation Center of Advanced Microstructures, Nanjing University, Nanjing, China}
\affiliation{MatricTime Digital Technology Co. Ltd., Nanjing, China}

\date{\today}

\begin{abstract}
Numerical methods are widely used to calculate the secure key rate of many quantum key distribution protocols in practice, but they consume many computing resources and are too time-consuming. In this work, we take the homodyne detection discrete-modulated continuous-variable quantum key distribution (CV-QKD) as an example, and construct a neural network that can quickly predict the secure key rate based on the experimental parameters and experimental results. Compared to traditional numerical methods, the speed of the neural network is improved by several orders of magnitude. Importantly, the predicted key rates are not only highly accurate but also highly likely to be secure. This allows the secure key rate of discrete-modulated CV-QKD to be extracted in real time on a low-power platform. Furthermore, our method is versatile and can be extended to quickly calculate the complex secure key rates of various other unstructured quantum key distribution protocols.
\end{abstract}

\maketitle

With the concurrent rise of artificial intelligence and quantum information science, these two fields are merging in a synergistic manner. In this growing trend, some works try to design new theoretical models based on quantum algorithms to improve classical machine learning for desired quantum speed-up \cite{lloyd2014quantum,ciliberto2018quantum,beer2020training,bondarenko2020quantum,farhi2018classification,mitarai2018quantum,wan2017quantum,chen2018quantum,jerbi2021quantum,abbas2021power}. At the same time, with the ever-increasing complexity of quantum systems, advanced quantum information technologies also require powerful tools for data processing and data analysis. We therefore urgently need to leverage existing classical machine learning techniques to solve practical, but difficult, problems in quantum information science, such as tomography \cite{torlai2018neural,smith2021efficient,quek2021adaptive}, classifying quantum states \cite{gao2018experimental,ma2018transforming,yang2019experimental}, quantum metrology \cite{hentschel2011efficient,fiderer2021neural,cimini2021calibration}, quantum control \cite{bukov2018reinforcement,wise2021using} and quantum cryptography\cite{coyle2020variational}.

Quantum key distribution (QKD)~\cite{bennett1984Quantum,ekert1991quantum} is by far the most practical technology in quantum information. It allows two distant parties (Alice and Bob) to establish secure keys against any eavesdropper. Various QKD protocols have been proposed one after another in recent decades \cite{xu2020secure,PRXQuantum.3.020315,yin2019phase,yin2020experimental,tang2021polarization,cui2019measurement}. Calculating the secure key rates of these QKD protocols is typically done by analytical methods \cite{matsuura2021finite}, but these analytical methods are usually inseparable from certain symmetry assumptions. These assumptions are often broken by experimental imperfections in practice. Therefore, to analyze the security of QKD protocols that are more suitable for practical implementations, some numerical methods based on convex optimization \cite{coles2016numerical,winick2018reliable,primaatmaja2019versatile,tan2021computing} have been developed.

For instance, continuous-variable (CV) QKD has its own distinct advantages at a metropolitan distance \cite{pirandola2020advances,zhang2020long} due to the use of common components of coherent optical communication technology. In addition, the homodyne \cite{grosshans2002continuous} or heterodyne \cite{weedbrook2004quantum} measurements used by CV-QKD have inherent extraordinary spectral filtering capabilities, which allows the crosstalk in wavelength division multiplexing (WDM) channels to be effectively suppressed. Therefore, hundreds of QKD channels may be integrated into a single optical fiber and can be cotransmitted with classic data channels. This allows QKD channels to be more effectively integrated into existing communication networks. In CV-QKD, discrete modulation technology has attracted much attention \cite{zhao2009asymptotic,leverrier2009unconditional,hirano2017implementation,ghorai2019asymptotic,lin2019asymptotic,lin2020trusted,liu2021homodyne,upadhyaya2021dimension,kanitschar2021tight,matsuura2021finite,kaur2021asymptotic,denys2021explicit} because of its ability to reduce the requirements for modulation devices. However, due to the lack of symmetry, the security proof of discrete modulation CV-QKD also mainly relies on numerical methods \cite{ghorai2019asymptotic,lin2019asymptotic,lin2020trusted,upadhyaya2021dimension,liu2021homodyne,kanitschar2021tight,hu2021robust}.

Unfortunately, calculating a secure key rate by numerical methods requires minimizing a convex function over all eavesdropping attacks related with the experimental data \cite{bunandar2020numerical,george2021numerical}. The efficiency of this optimization depends on the number of parameters of the QKD protocol. For example, in discrete modulation CV-QKD, the number of parameters is generally $1000-3000$ depending on the different choices of cutoff photon numbers~\cite{lin2019asymptotic}. This leads to the corresponding optimization possibly taking minutes or even hours \cite{hu2021robust}. Therefore, it is especially important to develop tools for calculating the key rate that are more efficient than numerical methods.

In this work, we take the homodyne detection discrete-modulated CV-QKD~\cite{lin2019asymptotic} as an example to construct a neural network capable of predicting the secure key rate for the purpose of saving time and resource consumption. We apply our neural network to a test set obtained at different excess noises and distances. Excellent accuracy and time savings are observed after adjusting the hyperparameters. Importantly, the predicted key rates are highly likely to be secure. Note that our method is versatile and can be extended to quickly calculate the complex secure key rates of various other unstructured quantum key distribution protocols. Through some open source deep learning frameworks for on-device inference, such as TensorFlow Lite \cite{tensorflow2015-whitepaper}, our model can also be easily deployed on devices at the \emph{edge} of the network, such as mobile devices, embedded Linux or microcontrollers.

\section*{Results}

\noindent
\textbf{Discrete-modulated CV-QKD.}
To clearly show the problem we try to solve, we briefly introduce the main ideas of discrete-modulated CV-QKD and give the convex optimization problem of finding its key rates in this section. See Ref. \cite{lin2019asymptotic} and Appendix A for a detailed description of discrete-modulated CV-QKD.

The protocol involves two parties, Alice and Bob. Alice randomly prepares one of the four coherent states and sends it to Bob by an untrusted quantum channel. Bob measures the received coherent state using homodyne detection. After repeating $N$ rounds, Alice and Bob perform sifting, parameter estimation, error correction and privacy amplification over the classical authentication channel to obtain the final secure key rates. The key rate formula in the asymptotic limit can be expressed according to Refs. \cite{winick2018reliable,coles2016numerical} as

\begin{equation}
R^{\infty}=\min _{\rho_{A B} \in \mathbf{S}} D\left(\mathcal{G}\left(\rho_{A B}\right) \| \mathcal{Z}\left[\mathcal{G}\left(\rho_{A B}\right)\right]\right)-p_{\mathrm{pass}} \delta_{\mathrm{EC}},
\end{equation}where $D(\rho \| \sigma)=\operatorname{Tr}\left(\rho \log _{2} \rho\right)-\operatorname{Tr}\left(\rho \log _{2} \sigma\right)$ is the quantum relative entropy; $\rho_{AB}$ is the bipartite state of Alice and Bob; $\mathcal{G}$ is the mapping to describe the postprocessing of the bipartite state $\rho_{A B}$; $\mathcal{Z}$ is a pinching quantum channel for reading out the results of the key rate mapping; $\mathbf{S}$ is the set of all density operators that match the experimental observations; $p_{\mathrm{pass}}$ is a sifting factor that determines how many rounds of data are used for generating keys; $\delta_{\mathrm{EC}}$ represents the amount of information leakage per bit in the error-correction process.

The key to finding the secure key rates is to solve the minimum value of $D\left(\mathcal{G}\left(\rho_{A B}\right) \| \mathcal{Z}\left[\mathcal{G}\left(\rho_{A B}\right)\right]\right)$, since $p_{\mathrm{pass}} \delta_{\mathrm{EC}}$ is a fixed quantity. The associated optimization problem is~\cite{lin2019asymptotic}

\begin{equation}\label{op}
\begin{split}
\operatorname{minimize}  &D\left(\mathcal{G}\left(\rho_{A B}\right) \| \mathcal{Z}\left[\mathcal{G}\left(\rho_{A B}\right)\right]\right) \\
\text{subject to } &\\
&\operatorname{Tr}\left[\rho_{A B}\left(|x\rangle\left\langle\left. x\right|_{A} \otimes \hat{q}\right)\right]=p_{x}\langle\hat{q}\rangle_{x}\right., \\
&\operatorname{Tr}\left[\rho_{A B}\left(|x\rangle\left\langle\left. x\right|_{A} \otimes \hat{p}\right)\right]=p_{x}\langle\hat{p}\rangle_{x}\right., \\
&\operatorname{Tr}\left[\rho_{A B}\left(|x\rangle\left\langle\left. x\right|_{A} \otimes \hat{n}\right)\right]=p_{x}\langle\hat{n}\rangle_{x}\right., \\
&\operatorname{Tr}[\rho_{A B}(|x\rangle \left\langle x\right|_{A} \otimes \hat{d})]=p_{x}\langle\hat{d}\rangle_{x}, \\
&\operatorname{Tr}\left[\rho_{A B}\right]=1, \\
&\rho_{A B} \geq 0,\\
&\operatorname{Tr}_{B}\left[\rho_{A B}\right]=\sum_{i, j=0}^{3} \sqrt{p_{i} p_{j}}\left\langle\varphi_{j} \mid \varphi_{i}\right\rangle|i\rangle\left\langle\left. j\right|_{A}\right.,
\end{split}
\end{equation}where $|x\rangle\left\langle\left. x\right|_{A}\right.$ is a local projective measurement operator of Alice's side, where $x \in\{0,1,2,3\}$; $\hat{q}=\frac{1}{\sqrt{2}}\left(\hat{a}^{\dagger}+\hat{a}\right)$, where $\hat{a}$ and $\hat{a}^{\dagger}$ are the annihilation and creation operators of a single-mode state, respectively; $\hat{p}=\frac{i}{\sqrt{2}}\left(\hat{a}^{\dagger}-\hat{a}\right)$; $\hat{n}=\frac{1}{2}\left(\hat{q}^{2}+\hat{p}^{2}-1\right)=\hat{a}^{\dagger} \hat{a}$; $\hat{d}=\hat{q}^{2}-\hat{p}^{2}=\hat{a}^{2}+\left(\hat{a}^{\dagger}\right)^{2}$; $\langle\hat{q}\rangle_{x}$, $\langle\hat{p}\rangle_{x}$, $\langle\hat{n}\rangle_{x}$ and $\langle\hat{d}\rangle_{x}$ represent the corresponding expectation values of the operators $\hat{q}$, $\hat{p}$, $\hat{n}$ and $\hat{d}$ acting on $\rho_{B}^{x}$, respectively; $\rho_{B}^{x}=\frac{1}{p_{x}} \operatorname{Tr}_{A}\left[\rho_{A B}\left(|x\rangle\left\langle\left. x\right|_{A} \otimes \mathrm{id}_{B}\right)\right]\right.$ is the state of Bob after Alice has performed measurement $|x\rangle\langle x|$ on $\rho_{A B}$, and $p_x$ is the corresponding probability; $\mathrm{id}_{B}$ is the identity transformation acting on system $B$.

The first four constraints in Eq.~(\ref{op}) are derived from experimental observations. The fifth and sixth constraints are conditions that the density matrix must satisfy. The seventh constraint comes from the fact that Alice's states do not change because they do not go through insecure quantum channels.

The optimization problem in Eq.~(\ref{op}) is to find the optimal $\rho_{A B}$ in $\mathbf{S}$ such that $R^{\infty}$ is minimized. $\rho_{A B}$ is infinite-dimensional because the attacker has the ability to arbitrarily perturb the optical mode sent by Alice into an infinite-dimensional state to send to Bob. To solve this optimization problem using numerical methods, we need to apply the photon-number cutoff assumption to $\rho_{A B}$ to ensure that the number of variables is in a reasonable range. A detailed description of this method can be found in Ref. \cite{lin2019asymptotic}.

After applying the photon-number cutoff assumption, the optimization problem in Eq.~(\ref{op}) can be solved by applying the numerical method in Refs. \cite{lin2019asymptotic,winick2018reliable}, but this is very time consuming. In this work, to reduce the time to predict secure key rates, we use the key rates obtained by the numerical method in Refs. \cite{lin2019asymptotic,winick2018reliable} as labels to train our neural network.

\bigskip
\noindent
\textbf{Neural networks for predicting the key rates.}
We use an artificial neural network to predict the key rates of discrete-modulated CV-QKD. The general spirit of the work is to encode the optimization problem in Eq.~(\ref{op}) on the loss function of a feedforward neural network and train the neural network by minimizing this loss function. The trained neural network can be seen as a mapping, which has learned the structure of the training set. For new instances, the neural network outputs the results directly via mapping, unlike traditional numerical methods that perform complex searches. As a result, the trained neural network saves a great deal of time, while ensuring a good level of accuracy. A more detailed description of neural networks can be found in Ref. \cite{goodfellow2016deep}.

\begin{figure*}[t]
  \centering
  \includegraphics[width=12cm]{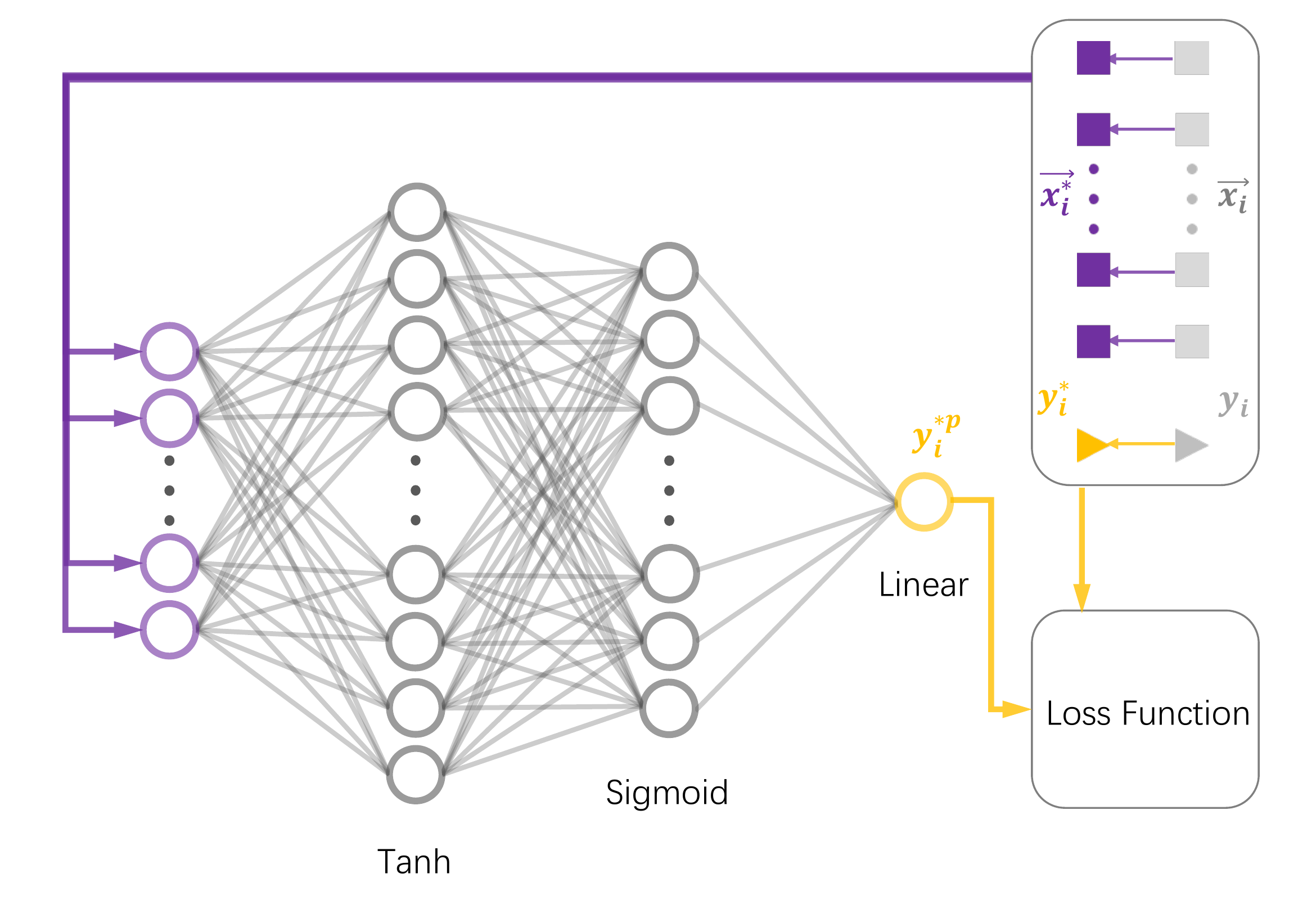}
\caption{Schematic diagram of our neural network model. We preprocess each training input $\vec{x}_i$ and its corresponding label $y_i$ to obtain $\vec{x}_i^*$ and $y_i^*$. The neural network receives $\vec{x}_i^*$ and outputs the corresponding $y_i^{*p}$. The numbers of neurons in the first hidden layer and the second hidden layer of the neural network are $400$ and $200$, respectively. $y_i^{*p}$ and $y_i^*$ are used to compute the loss function designed by us. Minimization of the loss function completes the training process.}
\label{fig1}
\end{figure*}

A four-layer neural network model is designed to predict the key rates of discrete-modulated CV-QKD (Fig.~\ref{fig1}). The input layer of the network has $29$ neurons, which are used to receive the training inputs. The first hidden layer and the second hidden layer of the network have $400$ and $200$ neurons respectively, and their activation functions are the tanh function and sigmoid function, respectively. The output layer has only one neuron, which is used to predict secure key rates.

\begin{figure*}[ht]
  \centering
  \includegraphics[width=18cm]{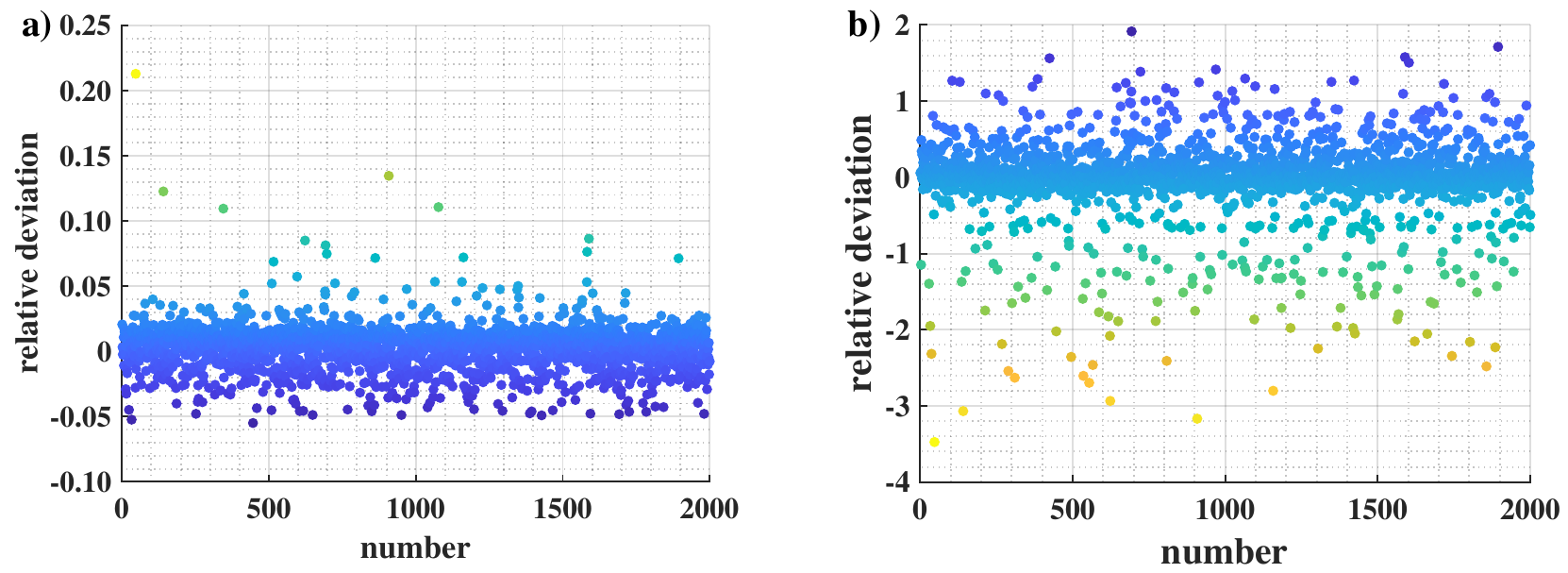}
	\caption{Relative deviations before and after data preprocessing. We use the network structure shown in Fig.~\ref{fig1} with the mean square error as the loss function to compare the results of data preprocessing (a) and without data preprocessing (b). The data set is generated under the excess noise of $0.002-0.005$, and is split into a training set containing $158000$ samples and a test set containing $2000$ samples. The horizontal coordinate represents the different samples in the test set. The vertical coordinate represents the relative deviations between the key rate predicted by our neural network and the key rate obtained by the numerical method at each sample.}
	\label{fig2}
\end{figure*}

To train our neural network, we generate the data set containing $552,000$ input instances $\left\{\vec{x}_{i}\right\}$ and $552,000$ corresponding labels $\left\{{y}_{i}\right\}$ using the numerical method in Refs. \cite{lin2019asymptotic,winick2018reliable}. Each $\vec{x}_{i} \in\left\{\vec{x}_{i}\right\}$ represents a vector of $29$ variables, and label ${y}_{i}$ represents the corresponding key rate. There are 16 variables in each $\vec{x}_{i}$ that are the right parts of the first four restrictions of Eq.~(\ref{op}), 12 variables in each $\vec{x}_{i}$ are nondiagonal elements of the right side matrix of the last restriction of Eq.~(\ref{op}), and the remaining variable is excess noise $\xi$. The 29 variables in each $\vec{x}_i$ can be calculated in the experiment by using experimental parameters and experimental observations. In our simulation, these random input instances $\left\{\vec{x}_{i}\right\}$ are generated directly from seven experimental parameters (transmission distance $L$, light intensity $\mu$, excess noise $\xi$, and probability $p0$, $p1$, $p2$ and $p3$) and the following method.

When the excess noise $\xi$ is within $0.002-0.014$, we first generate a two-dimensional grid with excess noise and distance in the horizontal and vertical coordinates, respectively. Specifically, the value of the distance is between $0$ and $100$ km in a step of $5$ km. The value of the excess noise is between $0.002$ and $0.014$ in a step of $0.001$. Then, each grid point is sampled $80$ times. With each sampling, the excess noise fluctuates around the exact value, and the float range is $0.0005$ up and down. Once the excess noise for this sampling is determined, the light intensity will take a value every $0.01$ between $0.35$ and $0.60$. Each sampling needs to generate $25$ input instances with a positive key rate; otherwise, the current round of sampling is discarded and restarted. In this way, $2000$ input instances are generated on each grid point. Correspondingly, a total of $520,000$ training inputs are generated on this two-dimensional grid. When the excess noise $\xi$ is $0.015$, a similar two-dimensional grid is generated. However, we only sample to $80$ km, so only $32,000$ instances are generated. In this way, we collect a total of $552,000$ samples with excess noise $\xi$ between $0.002$ and $0.015$. Using the numerical approach in Refs. \cite{lin2019asymptotic,winick2018reliable}, we calculate the corresponding key rate for each sample as the label of the data set on the blade cluster system of the High Performance Computing Center of Nanjing University. We consume over $40,000$ core hours, and the node we used contains 4 Intel Xeon Gold 6248 CPUs, which involves immense computational power.

To improve the convergence speed and accuracy of our neural network, we preprocess the input instances $\left\{\vec{x}_{i}\right\}$ and the corresponding labels $\left\{{y}_{i}\right\}$. To demonstrate the necessity of the data preprocessing, we use the network structure shown in Fig.~\ref{fig1} to perform a controlled experiment with the mean square error as the loss function. With the excess noise of $0.002-0.005$, the absolute values of the relative deviations between the key rates predicted by our neural network and the corresponding key rates obtained by the numerical method do not exceed $25\%$ after the data preprocessing (Fig.~\ref{fig2}), whereas the absolute values of the relative deviations exceed $400\%$ without the data preprocessing.  Here, the relative deviation is the absolute deviation between the predicted value and true value divided by the true value. A detailed description of the data preprocessing can be found in Appendix B.

A new loss function is specifically designed to make key rates predicted by our neural network as information-theoretically secure as possible, rather than using the traditional mean squared error as a loss function. The expression of the loss function is as follows:

\begin{align}\label{cost}
\begin{split}
C&=\frac{1}{n} \sum_{i=1}^{n} \gamma\left(e_{i}^{* 2}+\max \left(e_{i}^{*},-\log _{10}(\varepsilon)\right)\right)-(1-\gamma)\left(\min \left(e_{i}^{*}, 0\right)\right)
\end{split}
\end{align}, where $n$ is the number of training inputs. $e_i^*=y_{i}^{*p}-y_{i}^{ *}$ is the residual error between the preprocessed label $y_{i}^{*}$ and the corresponding output $y_{i}^{* p}$ of the neural network.

The minimum function part in Eq.~(\ref{cost}) is the penalty term and is used to make the key rates predicted by the neural network as information-theoretically secure as possible. On the other hand, the part consisting of the maximum function and the squared term in Eq.~(\ref{cost}) is used to bound the upper limit of $e_i^*$ to obtain higher key rates. The parameter $\gamma$ is used to balance the effects of the two parts. With the help of this loss function, we expect that the relative deviations between predicted value and true value can be bound in $(\varepsilon-1,0)$ after choosing the proper $\varepsilon$ and $\gamma$.

\begin{figure*}[ht]
  \centering
  \includegraphics[width=18cm]{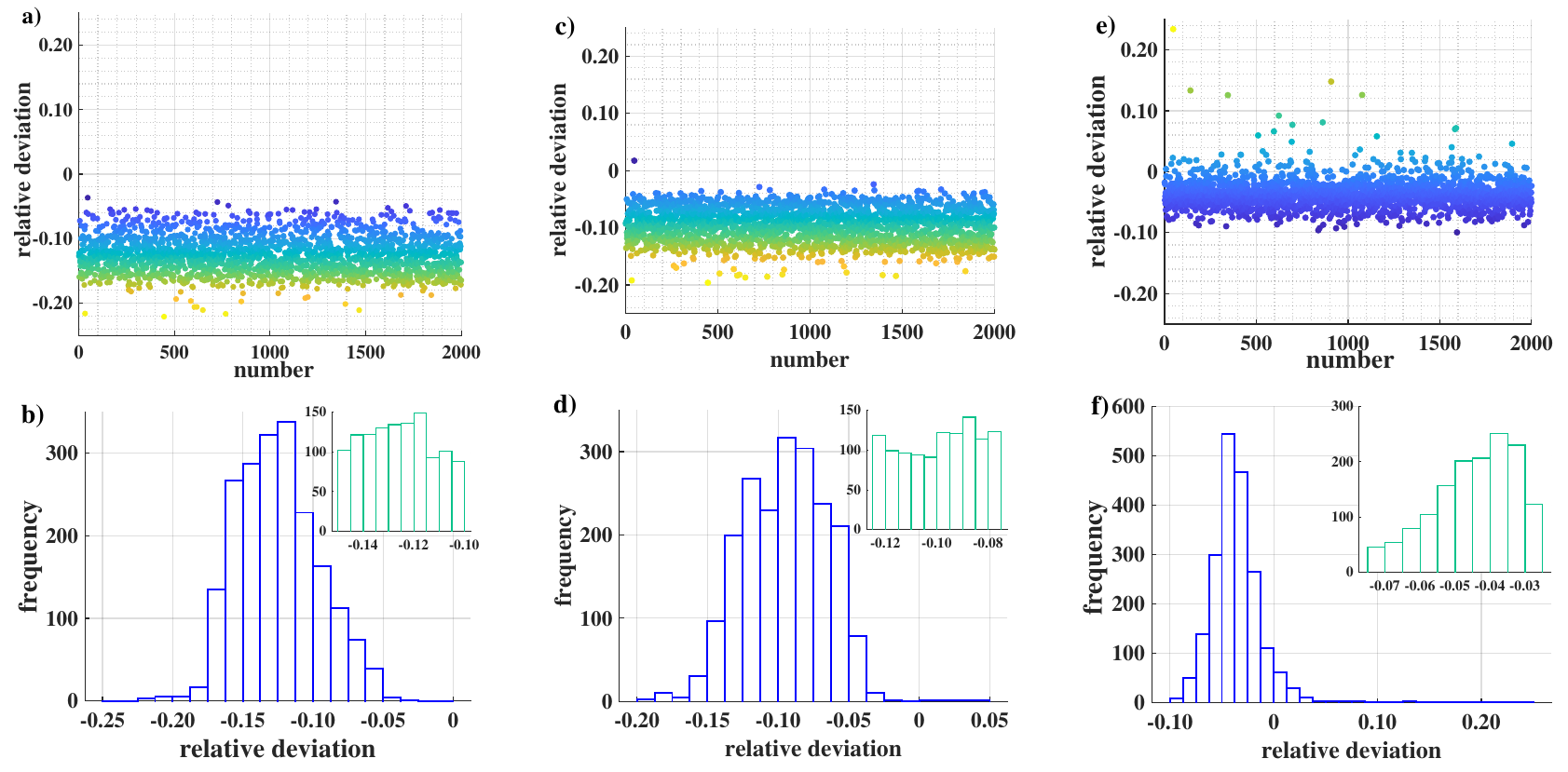}
	\caption{Performance comparison of neural networks with different hyperparameters. (a) shows the results of the neural network with the hyperparameters $\gamma=0.20$ and $\epsilon=0.80$ in predicting $2,000$ samples with excess noise between $0.002$ and $0.005$ in the test set. The predicted key rates are strictly below the key rates obtained by the numerical method in Refs. \cite{lin2019asymptotic,winick2018reliable}. (b) plots the histogram of the relative deviation distribution in (a). The absolute value of the relative deviations remains roughly in the region of $5\%-20\%$. (c-f) plot the corresponding results for the hyperparameters $\gamma=0.20$, $\epsilon=0.90$ and $\gamma=0.80$, $\epsilon=0.80$, respectively.}
	\label{fig3}
\end{figure*}

The performance of the neural networks is related to hyperparameters $\gamma$ and $\epsilon$. Without loss of generality, we take the examples of neural networks with excess noise $\xi$ between $0.002$ and $0.005$ (Fig.~\ref{fig3}). When $\gamma=0.20$ and $\epsilon=0.80$, the key rates predicted by the neural network are strictly lower than those obtained by the numerical method in Refs. \cite{lin2019asymptotic,winick2018reliable}, which means that the key rates predicted by the neural network are information-theoretically secure. Meanwhile, the absolute values of the relative deviations are mainly distributed between $0.05$ and $0.20$ (Fig.~\ref{fig3}a-b). Fig.~\ref{fig3}c-f plot the corresponding results for the hyperparameters $\gamma=0.20$, $\epsilon=0.90$ and $\gamma=0.80$, $\epsilon=0.80$, respectively. Note that the partial key rates predicted by the neural networks under $\gamma=0.20$, $\epsilon=0.90$ and $\gamma=0.80$, $\epsilon=0.80$ are higher than the key rates obtained by the numerical method. This indicates that the performance of neural networks trained with hyperparameters $\gamma=0.20$, $\epsilon=0.90$ and $\gamma=0.80$, $\epsilon=0.80$ is not as good as that of neural network trained with hyperparameters $\gamma=0.20$ and $\epsilon=0.80$. Therefore, we need to carefully tune hyperparameters of the neural networks to ensure their stable performance.

The $552,000$ data generated by the numerical method are split into a training set containing $524,400$ data and a test set containing $27,600$ data. The test set is sampled from the original data set and covers instances generated under all combinations of excess noise and distance. The data preprocessing procedure follows data splitting. The Adam optimization algorithm \cite{kingma2014adam} is used to train our neural network. The initial learning rate is set to $0.001$. For each training, we set $200$ epochs and $256$ batch sizes. In addition, techniques such as early stopping and dropout \cite{2014Dropout} are used to prevent overﬁtting. The relative deviations of the trained network on the test set and the training set have similar distributions, which indicates that the model has good generalization performance.

\begin{figure*}[ht]
  \centering
  \includegraphics[width=12cm]{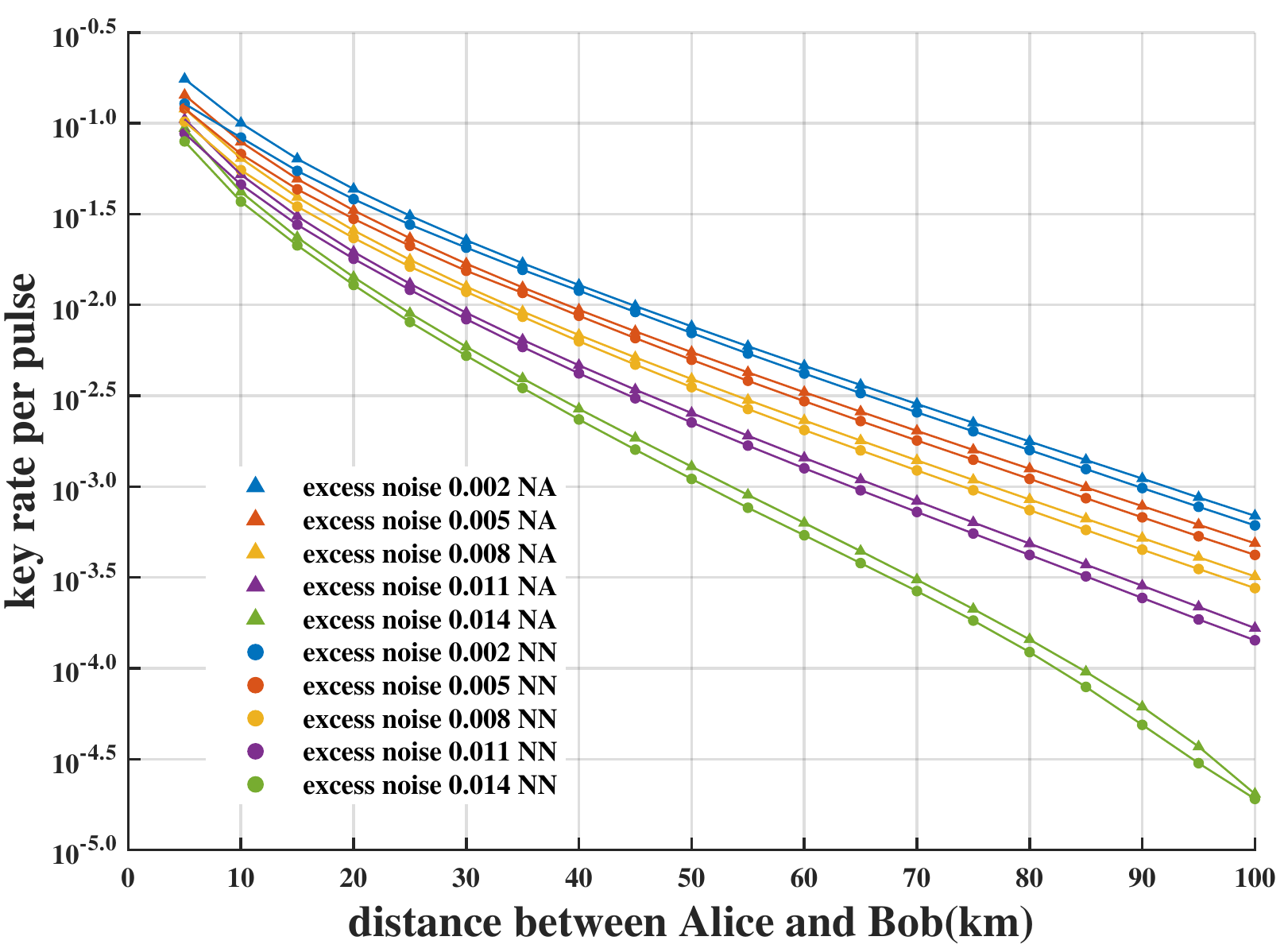}
\caption{Secure key rate versus the transmission distance for homodyne detection discrete-modulated CV-QKD with excess noise $\xi$ of $0.002$, $0.004$, $0.008$, $0.011$ and $0.014$ using our neural network (circles) and the numerical method in Refs. \cite{lin2019asymptotic,winick2018reliable} (triangles). The light intensity is chosen to be optimal in the interval $[0.35,0.6]$. Tht transmission efficiency $\eta=10^{-0.02 L}$. The reconciliation efficiency $\beta=0.95$. The neural network used for comparison is trained by setting the hyperparameters $\gamma=0.20$ and $\epsilon=0.80$. The cutoff photon number in the numerical method is set as $10$.}
\label{fig4}
\end{figure*}

\bigskip
\noindent
\textbf{Key rate comparison.}
We use our neural network to predict, given the optimal light intensity, key rates of discrete-modulated CV-QKD at different distances and different excess noises after training the neural network under $\gamma=0.20$ and $\epsilon=0.80$ according to the method described in Section III above. As shown in Fig.~\ref{fig4}, we compare the key rates with the corresponding key rates obtained by the numerical method in Refs. \cite{lin2019asymptotic,winick2018reliable}. The results show that all key rates predicted by the neural network are strictly lower than those obtained by the numerical method. It is worth noting that the relative deviations between them are basically within $20\%$ (relevant data can be found in Appendix C).

To illustrate the more general case, we test the test set containing $27,600$ samples mentioned at the end of Section III. The results show that the number of samples, for which the key rates predicted by the neural network are lower than the corresponding results calculated by the numerical method, is $27379$. Namely, the probability that the key rate predicted by the neural network on the test set is secure is as high as $99.2\%$.

\begin{figure*}[ht]
  \centering
  \includegraphics[width=12cm]{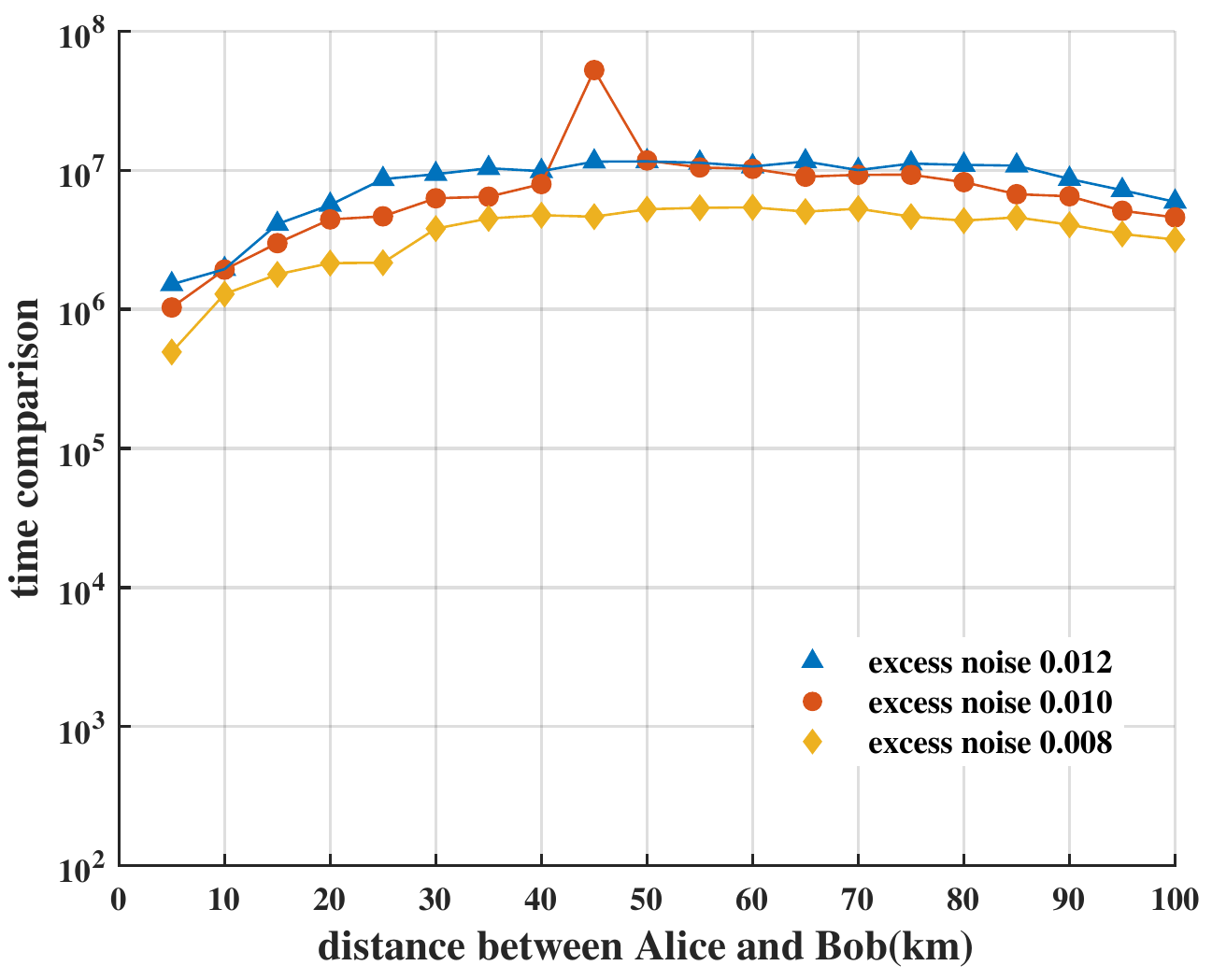}
\caption{Time consumption comparison between the neural network method and numerical method. The comparison results with excess noise of $0.008$, $0.010$ and $0.012$ are shown as diamonds, circles and triangles, respectively. Each point represents the logarithm of the ratio of the running time of the numerical method divided by the running time of the neural network method. The neural network used for comparison is trained by setting the hyperparameters $\gamma=0.20$ and $\epsilon=0.80$. The cutoff photon number  in the numerical method is set as $10$.}
\label{fig5}
\end{figure*}

Our neural network shows greater advantages over the numerical method in terms of time and resource consumption. We compare the time required to predict the key rates with our neural network and the time required to calculate the key rates with the numerical method on a high-performance personal computer with a 3.3 GHz AMD Ryzen 9 4900H and 16 GB of RAM (Fig.~\ref{fig5}). The neural network is $6-8$ orders of magnitude of the numerical method for predicting the key rates of the discrete-modulated CV-QKD within $0-100$ km for excess noise $\xi=0.008-0.012$. In addition, as the excess noise increases, the speed of the neural network increases even more. Refer to Appendix C for more detailed data.

\section*{Discussion}
We have constructed neural networks and shown that these neural networks can predict the information-theoretically secure key rates of homodyne detection discrete-modulated CV-QKD with a great probability (up to $99.2\%$) at a distance of $0-100$ km and an excess noise of no more than $0.015$. In particular, with excess noise up to $0.008$ or more, the speed of our method is at least improved by six orders of magnitude compared to that of the numerical method in Refs. \cite{lin2019asymptotic,winick2018reliable}. For example, it takes an average of $190$ seconds to numerically calculate the point with the excess noise $\xi$ around $0.008$, which greatly affects the efficiency of QKD systems to calculate the secure key rate. In contrast, a neural network can calculate tens of thousands of key rates in one second. Considering that it takes a certain amount of time for the QKD system to collect data, the speed of predicting the key rates by the neural network completely meets practical applications. This advantage brings us one step closer to achieving low latency for discrete modulated CV-QKD on a low-power platform. Our method is applicable in principle to any protocol that already has reliable numerical methods. However, for protocols such as $16/64/256$ QAM DM-CVQKD protocol with analytical methods whose effects are very close to those of numerical methods, it is not necessary to use the method proposed in this paper.

Recently, there have been two main types of situations in which machine learning is used in QKD. One is used for experimental parameter optimization~\cite{lu2019parameter,wang2019machine} and the other is used to assist experimental control~\cite{Liu2018Integrating,Liu2019Practical,chin2021machine}. They all use machine learning to replace traditional optimization or feedback control algorithms, which are significantly different from our work. To the best of our knowledge, this is the first time we have tried to apply machine learning methods to predict key rates of QKD. This poses a greater challenge than parameter optimization with machine learning methods. This is because the parameters predicted by the neural networks are substituted into numerical or analytical methods to find the corresponding key rates, which naturally ensures that the key rates are information-theoretically secure. However, the key rates obtained by neural networks do not guarantee this naturally, which forces us to redesign the loss function and seek better data preprocessing methods to guarantee the acquired key rate with information-theoretic security. Note that the probability ($0.8\%$) of our neural network predicting an insecure key rate is too large compared to conventional security parameters of the QKD protocol (e.g. $10^{-6}$). In practice, however, we need to sample thousands of data points and calculate their respective key rates to obtain a usable keystring. The key here is that when we sum and average the key rates of all data points predicted by our neural network, the insecure probability of this averaged key rate can be reduced very low. If there are enough data points, this insecure probability can also approximate conventional security parameters of the QKD protocol.

We expect that larger excess noises and longer distances will require a deeper network, more sophisticated loss functions, and more detailed data preprocessing methods to improve the performance of neural networks on the training set. More training data are also necessary to improve the generalization ability of the neural networks. For deep neural networks, the rapid growth or rapid disappearance of the transmitted gradient hinders the optimization process; therefore, the debugging process is highly technical. The debugging process can be guided by monitoring the activation function values of the neurons and histograms 1 of those gradients \cite{goodfellow2016deep}.

Our machine learning approach is at least six orders of magnitude of the numerical method at predicting the secure key rates of homodyne detection discrete-modulated CV-QKD with excess noise up to $0.008$ or more. However, training our neural network is still time consuming. This is because we need to use traditional numerical methods to obtain a number of key rates as the training set of the neural networks. In particular, the performance of our neural network is dependent on the choice of hyperparameters $\gamma$, $\epsilon$ and initial learning rate. This means that we may need to train several times to obtain a suitable neural network. To make our machine learning method more intelligent, further work is necessary to design another neural network to automatically find the most suitable hyperparameters. We have also tried other machine learning methods, such as boosting decision trees. These methods have smaller relative deviations, but have greater variances. We have left the fusion of these methods to future research.

The important contribution of our work is that it opens the door to using classical machine learning to predict QKD key rates. In particular, our ideas and methods are very easy to generalize to other QKD protocols. We expect that our work will stimulate further research to help most QKD systems run on low-power chips \cite{kwek2021chip} in mobile devices \cite{wang2021transmission}.

\section*{Methods}
\noindent
\textbf{Discrete-modulated CV-QKD.}
According to Ref. \cite{lin2019asymptotic}, homodyne detection discrete-modulated CV-QKD is described below:

(1) State preparation.—Alice prepares a coherent state $\left|\psi_k\right\rangle$ from the set $ \{|\alpha\rangle,|-\alpha\rangle,|i \alpha\rangle,|-i \alpha\rangle\} $ according to the probability of $ [p_A/2,p_A/2,(1-p_A)/2,(1-p_A)/2]$, where $\alpha \in R$ is a predetermined amplitude and $k$ is the number of rounds. Then Alice sends the state $\left|\psi_k\right\rangle$ to Bob.

(2) Measurement.—Bob performs a homodyne measurement on the received state. He chooses to measure a certain orthogonal component ($q$ or $p$) according to the probability of $ [p_B,1-p_B] $. If $q$ is chosen, Bob notes $ b_k=0 $, otherwise he notes $ b_k=1$. Then, Bob records his measurement outcome $y_{k} \in R$.

(3) Announcement and sifting.—After repeating the first two steps $N$ times, Alice and Bob communicate via the classical authentication channel and divide the obtained data into the following four subsets:

\begin{equation}
 \begin{aligned}
\mathcal{I}_{q q} &=\left\{k \in[N]:\left|\psi_{k}\right\rangle \in\{|\alpha\rangle,|-\alpha\rangle\}, b_{k}=0\right\}, \\
\mathcal{I}_{q p} &=\left\{k \in[N]:\left|\psi_{k}\right\rangle \in\{|\alpha\rangle,|-\alpha\rangle\}, b_{k}=1\right\}, \\
\mathcal{I}_{p q} &=\left\{k \in[N]:\left|\psi_{k}\right\rangle \in\{|i \alpha\rangle,|-i \alpha\rangle\}, b_{k}=0\right\}, \\
\mathcal{I}_{p p} &=\left\{k \in[N]:\left|\psi_{k}\right\rangle \in\{|i \alpha\rangle,|-i \alpha\rangle\}, b_{k}=1\right\},
\end{aligned}
\end{equation}where $[N]$ denotes the set of all integers from $1$ to $N$. Then Alice and Bob randomly select a subset $\mathcal{I}_{\text {key }}$ of size $m$ from $\mathcal{I}_{q q}$ for generating keys. The key string $\mathbf{X}=\left(x_{1}, x_{2}, \ldots, x_{m}\right)$ at Alice is also determined according to the following rules:

\begin{equation}
\forall j \in[m], \quad x_{j}=\left\{\begin{array}{ll}
0 & \text { if }\left|\psi_{f(j)}\right\rangle=|\alpha\rangle, \\
1 & \text { if }\left|\psi_{f(j)}\right\rangle=|-\alpha\rangle,
\end{array}\right.
\end{equation}where $f(j)$ is a function that maps from $\mathcal{I}_{\text {key }}$ to $\mathcal{I}_{q q}$. The remaining data in $\mathcal{I}_{q q}$, $\mathcal{I}_{q p}$, $\mathcal{I}_{p q}$ and $\mathcal{I}_{pp}$ are integrated into the set $\mathcal{I}_{\text {test }}$ and used for parameter estimation.

(4) Parameter estimation.—Alice and Bob perform parameter estimation based on the data in $\mathcal{I}_{\text {test }}$. First, they calculate the first and second moments of $q$ and $p$ quadratures for each of the four coherent states sent by Alice. Then they calculate the secret key rate based on the convex optimization problem in Eq. (8).

If the result shows that the key rate is equal to $0$, Alice and Bob abort the protocol and start over. Otherwise, they continue with the next step.

(5) Reverse reconciliation key map.—The key string $\mathbf{Z}=\left(z_{1}, z_{2}, \ldots, z_{m}\right)$ at Bob is determined according to Bob's measurement outcome $y_k$ in step 2 and the following rules:

\begin{equation}
z_{j}=\left\{\begin{array}{ll}
0 & \text { if } y_{f(j)} \in\left[\Delta_{c}, \infty\right), \\
1 & \text { if } y_{f(j)} \in\left(-\infty,-\Delta_{c}\right], \\
\perp & \text { if } y_{f(j)} \in\left(-\Delta_{c}, \Delta_{c}\right),
\end{array}\right.
\end{equation}where $\Delta_{c} \geq 0$ is determined by the postselection of data.

Alice and Bob then pick out the location of the symbol $\perp$ and remove the data at that location by classical communication. The set $\mathbf{X}$ and $\mathbf{Z}$ after removing $\perp$ is the raw key string.

(6) Error correction and privacy amplification.—Alice and Bob choose a suitable error-correction protocol and a suitable privacy-amplification protocol to generate secret key rates.

The key rate can be calculated using the well-known Devetak-Winter formula \cite{devetak2005distillation} in the asymptotic limit and under collective attacks. To apply this formula, we transform the prepare-and-measure protocol into the entanglement-based protocol.

Alice prepares the state according to the ensemble $\left\{\left|\varphi_{x}\right\rangle, p_{x}\right\}$ in the prepare-and-measure protocol. In the equivalent entanglement-based protocol, Alice prepares the  bipartite state in the form of $|\Psi\rangle_{A A^{\prime}}=\sum_{x} \sqrt{p_{x}}|x\rangle_{A}\left|\varphi_{x}\right\rangle_{A^{\prime}}$. Here Alice keeps $|x\rangle_{A}$ in register $A$ and sends $\left|\varphi_{x}\right\rangle_{A^{\prime}}$ to Bob. $\left|\varphi_{x}\right\rangle_{A^{\prime}}$ changes as it passes through an insecure quantum channel. The process can be described by a completely positive and trace-preserving map $\mathcal{E}_{A^{\prime} \rightarrow B}$. The bipartite state $\rho_{A B}$ thus transforms into

\begin{equation}
\rho_{A B}=\left(\mathrm{id}_{A} \otimes \mathcal{E}_{A^{\prime} \rightarrow B}\right)\left(|\Psi\rangle\left\langle\left.\Psi\right|_{A A^{\prime}}\right)\right.,
\end{equation}where $\mathrm{id}_{A}$ is the identity transformation acting on $A$. Under reverse reconciliation \cite{grosshans2003virtual}, the key rate formula can be expressed according to Refs. \cite{coles2016numerical,winick2018reliable} as

\begin{equation}
R^{\infty}=\min _{\rho_{A B} \in \mathbf{S}} D\left(\mathcal{G}\left(\rho_{A B}\right) \| \mathcal{Z}\left[\mathcal{G}\left(\rho_{A B}\right)\right]\right)-p_{\mathrm{pass}} \delta_{\mathrm{EC}}.
\end{equation}

\begin{algorithm}
    \caption{Training stage}
    \label{alg1}
    \DontPrintSemicolon
    \KwIn{ \ttt{\{$(\vec{x}_i, y_i)$\}} \tcp*[1]{Original training data set of discrete-modulated CV-QKD collected from the numerical method. $\vec{x_i}$ is feature vector containing 29 variables, and ${y_i}$ is the corresponding key rate.}}
    \KwIn{\ttt{$\gamma, \epsilon$} \tcp*[1]{Two hyperparameters in our self-designed loss function.}}
    \KwOut{${\{\theta_r\}}$ \tcp*[1]{The final learned weights of the neural network $\aleph_r$}}\;

    \ttt{Preprocessing } $\{\vec{x}_i^*\} \gets \{\vec{x}_i\}$ :

	\ttt{~~~Calculate the mean vector $\vec{x}$ of \{$\vec{x}_i$\}}

    \ttt{~~~Calculate the variance vector ${\vec{\sigma}}$ of \{$\vec{x}_i$\}}

	\uIf {${{\sigma_j}} == 0$}{
		$x_{ij}^*=x_{ij}$
	}
	\uElse{
		$x_{ij}^*=(x_{ij}-{\mu_j})/{\sigma_j}$
	}

    \ttt{Preprocessing } $\{y_i^*\} \gets \{y_i\}$:

	$y_i^*=-\log_{10}({y_i})$

    \ttt{Train the neural network under\{$\gamma, \epsilon$\} with ${\{(\vec x_i^*,y_i^*)\}}$}

    return ${\{\theta_r\}}$

\end{algorithm}

\begin{algorithm}
    \caption{Inference stage}
    \label{alg2}
    \DontPrintSemicolon
    \KwIn{ \ttt{\{$\vec{x}_i$\}} \tcp*[1]{A set of original feature vectors containing 29 variables collected from the experiment.}}
    \KwOut{$\{y_i\}$ \tcp*[1]{A set of corresponding key rates predicted by Neural network $\aleph_r$.}}\;

    \ttt{Preprocessing } $\{\vec{x}_i^*\} \gets \{\vec{x}_i\}$

    \For {$\vec{x}_i^* \in \{\vec{x}_i^*\}$
    }{
   	 $y_i^*$ $\gets$ $\aleph_r$($\vec{x_i^*}$)
   	
    	${y_i} = 10^{-y_i^*}$
    }

    return ${\{{y_i}\}}$

\end{algorithm}

\bigskip
\noindent
\textbf{Details of data preprocessing.}
To improve the performance of our neural network, we preprocess the training inputs $\left\{\vec{x}_{i}\right\}$ before training the neural network. The process can be expressed as

\begin{table*}[]
\caption{Relative deviations between key rates predicted by our neural network and the corresponding key rates obtained by the numerical method for the given optimal light intensity at different distances and different excess noises.}
\begin{tabular}{c|c|c|c|c|c}
\hline
\hline
\multicolumn{1}{c}{L}
&
\multicolumn{5}{|c}{Relative deviations}
\\
\multicolumn{1}{c}{(km)}
&
\multicolumn{1}{|c}{$\xi=0.002$}
&
\multicolumn{1}{|c}{$\xi=0.005$}
&
\multicolumn{1}{|c}{$\xi=0.008$}
&
\multicolumn{1}{|c}{$\xi=0.011$}
&
\multicolumn{1}{|c}{$\xi=0.014$}
\\
\hline
5& $0.27 $   & $0.16$  & $0.15$ & $0.16$  & $0.15$\\
10& $0.17 $   & $0.14$  & $0.14$ & $0.12$  & $0.12$\\
15& $0.14 $   & $0.12$  & $0.11$ & $0.10$  & $0.10$\\
20& $0.12 $   & $0.10$  & $0.09$ & $0.08$  & $0.09$\\
25& $0.11 $   & $0.09$  & $0.08$ & $0.07$  & $0.10$\\
30& $0.09 $   & $0.09$  & $0.06$ & $0.08$  & $0.11$\\
35& $0.08 $   & $0.07$  & $0.06$ & $0.09$  & $0.11$\\
40& $0.07 $   & $0.07$  & $0.08$ & $0.10$  & $0.13$\\
45& $0.07 $   & $0.08$  & $0.09$ & $0.10$  & $0.14$\\
50& $0.08 $   & $0.09$  & $0.10$ & $0.11$  & $0.14$\\
55& $0.09 $   & $0.10$  & $0.11$ & $0.12$  & $0.15$\\
60& $0.09 $   & $0.11$  & $0.11$ & $0.12$  & $0.14$\\
65& $0.10 $   & $0.11$  & $0.12$ & $0.13$  & $0.14$\\
70& $0.10 $   & $0.11$  & $0.12$ & $0.13$  & $0.13$\\
75& $0.10 $   & $0.12$  & $0.12$ & $0.13$  & $0.14$\\
80& $0.10 $   & $0.12$  & $0.13$ & $0.13$  & $0.15$\\
85& $0.11 $   & $0.13$  & $0.13$ & $0.14$  & $0.17$\\
90& $0.11 $   & $0.13$  & $0.14$ & $0.14$  & $0.20$\\
95& $0.11 $   & $0.14$  & $0.14$ & $0.15$  & $0.19$\\
100& $0.11 $   & $0.14$  & $0.14$ & $0.14$  & $0.06$\\
\hline
\hline
\end{tabular}
\label{table1}
\end{table*}

\begin{table*}[]

\caption{Time consumption of the neural network versus the numerical method with excess noise $\xi$ of $0.008$, $0.010$ and $0.012$. NM and NN are the abbreviations of the numerical method and neural network, respectively. L is the distance between Alice and Bob.}
\begin{tabular}{ccc|ccc|ccc}
\hline
\hline
\multicolumn{3}{c}{$\xi=0.008$}
&
\multicolumn{3}{|c}{$\xi=0.010$}
&
\multicolumn{3}{|c}{$\xi=0.012$}
\\
L(km) ~~~~~~& NM(s) ~~~~~~& NN(s)
~~~~~~& L(km) ~~~~~~& NM(s) ~~~~~~& NN(s) ~~~~~~& L(km)~~~~~~ & NM(s) ~~~~~~& NN(s)
\\
\hline
5 & $1.42\times 10^2 $  ~~~~~~& $1.98 \times 10^{-4}$ & 5 & $1.54 \times 10^2 $ ~~~~~~& $3.28 \times 10^{-4}$ & 5 & $2.16 \times 10^2$ ~~~~~~& $1.31 \times 10^{-4}$\\
10& $7.86 \times 10^1 $   ~~~~~~& $7.25 \times 10^{-5}$ & 10 & $9.94 \times 10^1 $  ~~~~~~& $5.85 \times 10^{-5}$ & 10 & $1.27 \times 10^2$  ~~~~~~& $4.70 \times 10^{-5}$\\
15& $1.04 \times 10^2 $   ~~~~~~& $6.60 \times 10^{-5}$ & 15 & $1.72 \times 10^2 $  ~~~~~~& $5.70 \times 10^{-5}$ & 15 & $2.24 \times 10^2$  ~~~~~~& $4.15 \times 10^{-5}$\\
20& $1.09 \times 10^2 $   ~~~~~~& $6.50 \times 10^{-5}$ & 20 & $2.37 \times 10^2 $  ~~~~~~& $5.40 \times 10^{-5}$ & 20 & $3.07 \times 10^2$  ~~~~~~& $4.30 \times 10^{-5}$\\
25& $1.20 \times 10^2 $   ~~~~~~& $6.65 \times 10^{-5}$ & 25 & $2.45 \times 10^2 $  ~~~~~~& $6.30 \times 10^{-5}$ & 25 & $4.40 \times 10^2$  ~~~~~~& $4.25 \times 10^{-5}$\\
30& $1.98 \times 10^2 $   ~~~~~~& $5.65 \times 10^{-5}$ & 30 & $3.30 \times 10^2 $  ~~~~~~& $4.75 \times 10^{-5}$ & 30 & $4.92 \times 10^2$  ~~~~~~& $4.20 \times 10^{-5}$\\
35& $2.34 \times 10^2 $   ~~~~~~& $5.90 \times 10^{-5}$ & 35 & $3.71 \times 10^2 $  ~~~~~~& $5.90 \times 10^{-5}$ & 35 & $5.33 \times 10^2$  ~~~~~~& $4.65 \times 10^{-5}$\\
40& $2.47 \times 10^2 $   ~~~~~~& $5.70 \times 10^{-5}$ & 40 & $4.18 \times 10^2 $  ~~~~~~& $5.85 \times 10^{-5}$ & 40 & $5.72 \times 10^2$  ~~~~~~& $4.60 \times 10^{-5}$\\
45& $2.50 \times 10^2 $   ~~~~~~& $6.10 \times 10^{-5}$ & 45 & $2.73 \times 10^2 $  ~~~~~~& $5.70 \times 10^{-5}$ & 45 & $5.94 \times 10^2$  ~~~~~~& $4.35 \times 10^{-5}$\\
50& $2.62 \times 10^2 $   ~~~~~~& $6.35 \times 10^{-5}$ & 50 & $6.24 \times 10^2 $  ~~~~~~& $5.60 \times 10^{-5}$ & 50 & $5.79 \times 10^2$  ~~~~~~& $4.55 \times 10^{-5}$\\
55& $2.74 \times 10^2 $   ~~~~~~& $6.50 \times 10^{-5}$ & 55 & $5.55 \times 10^2 $  ~~~~~~& $5.10 \times 10^{-5}$ & 55 & $5.83 \times 10^2$  ~~~~~~& $4.30 \times 10^{-5}$\\
60& $2.68 \times 10^2 $   ~~~~~~& $6.65 \times 10^{-5}$ & 60 & $5.28 \times 10^2 $  ~~~~~~& $5.85 \times 10^{-5}$ & 60 & $5.96 \times 10^2$  ~~~~~~& $4.30 \times 10^{-5}$\\
65& $2.55 \times 10^2 $   ~~~~~~& $6.70 \times 10^{-5}$ & 65 & $5.48 \times 10^2 $  ~~~~~~& $5.10 \times 10^{-5}$ & 65 & $5.96 \times 10^2$  ~~~~~~& $4.20 \times 10^{-5}$\\
70& $2.72 \times 10^2 $   ~~~~~~& $6.55 \times 10^{-5}$ & 70 & $4.82 \times 10^2 $  ~~~~~~& $5.65 \times 10^{-5}$ & 70 & $5.91 \times 10^2$  ~~~~~~& $5.30 \times 10^{-5}$\\
75& $2.60 \times 10^2 $   ~~~~~~& $6.70 \times 10^{-5}$ & 75 & $4.78 \times 10^2 $  ~~~~~~& $6.70 \times 10^{-5}$ & 75 & $5.87 \times 10^2$  ~~~~~~& $4.10 \times 10^{-5}$\\
80& $2.30 \times 10^2 $   ~~~~~~& $6.00 \times 10^{-5}$ & 80 & $4.19 \times 10^2 $  ~~~~~~& $5.20 \times 10^{-5}$ & 80 & $5.57 \times 10^2$  ~~~~~~& $4.35 \times 10^{-5}$\\
85& $2.34 \times 10^2 $   ~~~~~~& $5.70 \times 10^{-5}$ & 85 & $3.63 \times 10^2 $  ~~~~~~& $5.95 \times 10^{-5}$ & 85 & $5.45 \times 10^2$  ~~~~~~& $4.35 \times 10^{-5}$\\
90& $1.99 \times 10^2 $   ~~~~~~& $5.75 \times 10^{-5}$ & 90 & $3.48 \times 10^2 $  ~~~~~~& $5.35 \times 10^{-5}$ & 90 & $4.37 \times 10^2$  ~~~~~~& $4.10 \times 10^{-5}$\\
95& $1.72 \times 10^2 $   ~~~~~~& $5.75 \times 10^{-5}$ & 95 & $2.92 \times 10^2 $  ~~~~~~& $5.10 \times 10^{-5}$ & 95 & $3.81 \times 10^2$  ~~~~~~& $4.35 \times 10^{-5}$\\
100& $1.54 \times 10^2 $   ~~~~~~& $5.85 \times 10^{-5}$ & 100 & $2.43 \times 10^2 $  ~~~~~~& $6.60 \times 10^{-5}$ & 100 & $3.47 \times 10^2$  ~~~~~~& $4.65 \times 10^{-5}$\\
\hline
\hline
\end{tabular}
\label{table2}
\end{table*}

\begin{equation}\label{pre}
x_{i j}^{*}=\frac{x_{i j}-\bar{x}_{j}}{ \sigma_{j}},
\end{equation} where $x_{i j}$ represents the $j$-th component of the $i$-th sample; $\bar{x}_{j}$ and $\sigma_{j}$ are the mean and variance of the $j$-th component in all samples, respectively; $x_{i j}^{*}$ is the $j$-th component of the $i$-th sample after being preprocessed.

The preprocessed data $\{\vec{x}_i^*\}$ follow a standard normal distribution with a mean of $0$ and a variance of $1$. The process removes dimensional restrictions and facilitates the comparison of features of different dimensions. Since the maximum difference between different key rates in these samples is $4$ orders of magnitude, we preprocess the labels as follows to speed up the training process of the neural networks:

\begin{equation}
y_{i}^{ *}=-\log _{10}\left(y_{i}\right),
\end{equation} where $y_{i}^{ *}$ is the label corresponding to the $i$-th sample after being preprocessed. Note that the outputs predicted by the neural networks trained with preprocessed labels $\{y_{i}^{ *}\}$ need to be inverse solved using the following equation:

\begin{equation}
y_{i}^{ p}=10^{-y_{i}^{*p}},
\end{equation} where $y_{i}^{*p}$ and $y_{i}^{ p}$ are the output value and the predicted key rate of the neural networks for the $i$-th sample, respectively.

Algorithms ~\ref{alg1} and ~\ref{alg2} show the detailed training process of the neural networks and the process of using trained neural networks to predict new samples, respectively.

\bigskip
\noindent\textbf{Detailed data.}
Table ~\ref{table1} shows the relative deviations between the key rates predicted by our neural network and the corresponding key rates obtained by the numerical method for the given optimal light intensity at different distances and different excess noises. This table is a supplement to Fig.~\ref{fig4}.

Table ~\ref{table2} shows the specific data of the time consumption of the neural network and the numerical method with excess noise $\xi$ of $0.008$, $0.010$ and $0.012$. In the numerical method, each point with excess noise $\xi$ of approximately $0.01$ takes $200$ seconds on average, which greatly affects the efficiency of the QKD system to calculate the secure key rate. In contrast, the neural network can calculate tens of thousands of key rates in one second. Considering that it takes a certain amount of time for the QKD system to collect data, the speed of predicting the key rates by the neural network completely meets practical applications.

\bigskip
\noindent\textbf{Acknowledgements}\\
We gratefully acknowledge the support from the Natural Science Foundation of Jiangsu Province (No. BK20211145), the Fundamental Research Funds for the Central Universities (No. 020414380182), the Key Research and Development Program of Nanjing Jiangbei New Aera (No. ZDYD20210101), the Key-Area Research and Development Program of Guangdong Province (No. 2020B0303040001). We are grateful to the High Performance Computing Center of Nanjing University for performing the numerical calculations in this paper on its blade cluster system.


%

\end{document}